\documentclass[iop, apj, amsfonts, notitlepage, nofootinbib, twocolumn, preprintnumbers]{revtex4-1}
\usepackage{hyperref}
\usepackage{slashed}
\usepackage{natbib}
\usepackage{graphics,graphicx}
\hypersetup{
colorlinks=true,
linkcolor=red,
citecolor=blue,
urlcolor=blue}
\usepackage{epsf,epsfig}
\usepackage{amscd}
\usepackage{amsmath}

\input xy
\xyoption{all} 

\newcommand{\figref}[1]{Fig.~\ref{#1}}

\newcommand{\reference}[1]{Ref.~\cite{#1}}
\newcommand{\ourpreviouspaper}{\reference{Bedaque:2011hs} }

\newcommand{\oneover}[1]{\ensuremath{\frac{1}{#1}}}
\newcommand{\half}{\oneover{2}}
\newcommand{\inverse}{\ensuremath{^{-1}}}

\newcommand{\goesto}{\ensuremath{\rightarrow}}

\newcommand{\nn}{\nonumber}

\newcommand{\abs}[1]{\ensuremath{\left|	#1	\right|}}
\newcommand{\adj}{\ensuremath{^{\dagger}}}

\newcommand{\Lag}{{\mathcal L}}
\newcommand{\grad}{\ensuremath{\nabla}}

\newcommand{\Z}{\ensuremath{Z}}
\newcommand{\thw}{\ensuremath{\theta_W}}
\newcommand{\Gf}{\ensuremath{G_F}}

\newcommand{\Hefour}{\ensuremath{^4\text{He}}}
\newcommand{\sun}{\ensuremath{\odot}}

\begin{document}

\title{Neutrino Emission from Helium White Dwarfs with Condensed Cores} 
\author{Paulo F.~Bedaque}\email{{\tt bedaque@umd.edu}} 
\author{Evan Berkowitz}\email{{\tt evanb@umd.edu}}
\affiliation{Maryland Center for Fundamental Physics,\\ Department of Physics,\\ University of Maryland, College Park, MD USA} 
\author{Aleksey Cherman}\email{{\tt a.cherman@damtp.cam.ac.uk}}
\affiliation{Department of Applied Mathematics and Theoretical Physics,\\ University of Cambridge, Cambridge CB3 0WA, UK }

\preprint{UMD-DOE/40762-516}
\preprint{DAMTP-2012-20} 

\begin{abstract}
The possibility that ions in a helium white dwarf star are in a Bose-Einstein condensed state has been explored recently. In particular, it has been argued that the resulting novel quantum liquid has a new kind of quasiparticle excitation with a phonon-like dispersion relation. We investigate the neutrino emission rate due to this gapless state and the resulting impact on the total luminosity of helium white dwarf stars, as a possible observable way of detecting this exotic phase.  If the condensation temperature for the quantum liquid state, which is currently not known very precisely, turns out to be high enough, our calculations indicate that neutrino emission due to the gapless mode would make a large contribution to the total luminosity of the helium white dwarf stars. 
\end{abstract}
\maketitle

{\it Introduction.}
\label{sec:introduction}
Stars which are not too heavy become white dwarfs upon running out of fuel for nuclear fusion. 
White dwarf stars evolve by cooling, and have two cooling mechanisms.  Very young white dwarfs lose energy mostly through the decay of massive plasmon quasiparticles to neutrinos for temperatures $\gtrsim 10^{9}\,\mathrm{Kelvin}$ \cite{1969ApJ...155..221S,1966ApJ...146..437V,1969ApJ...156.1021K}. At these temperatures the effect of neutrino emission may be detected in frequency changes of non-radial oscillations of the star\cite{1985ApJ...298..752K,2000ApJ...539..372O}.
 However, plasmons become Boltzmann-suppressed for $T \lesssim 10^{9}\,\mathrm{Kelvin}$, and below this temperature the dominant energy loss mechanism for white dwarfs  is believed to be electromagnetic radiation from the surface of the star. 

Here we discuss the neutrino emission rate for helium-core white dwarfs (He WDs).  He WDs are much rarer than the usual carbon-oxygen white dwarfs (CO WDs), and are only  formed if a red-giant stage star in {\it e.g.} a binary system loses much of its hydrogen envelope before helium burning can begin.   The first He WDs candidates were detected only recently\cite{1999ApJ...516..250E,2001ApJ...553L.169T,2009ApJ...699...40S} in the globular cluster NGC 6397.  In \reference{2009ApJ...699...40S} it was observed that the He WD sequence terminates early compared to the CO WD sequence, so that there is an apparent dearth of faint He WDs.  This unexpected result motivates a reexamination of the physics of super-dense helium plasmas.

Recently it was realized that for low enough temperatures, the helium nuclei in a high density helium plasma should Bose-condense, forming a `nuclear condensate'\cite{Gabadadze:2008mx,Gabadadze:2009dz,Gabadadze:2009jb,Bedaque:2011hs}.  At the relevant temperatures and densities, the specific heat $c_{v}$ of the condensed phase turns out to be $\sim 10^{-2}$ times smaller than the $c_{v}$ of the classical plasma phase, and $\sim10^{-4}$ times smaller than the specific heat of an ion lattice, if one were formed.  This implies that He WDs with condensed core should cool by electromagnetic radiation significantly faster than WDs with uncondensed cores.  Whether this cooling speedup is large enough to be observable depends on detailed modeling\cite{Gabadadze:2009dz,Benvenuto:2011fj}, and the situation is currently not entirely clear.   

In \ourpreviouspaper we showed that the condensed phase of super-dense helium has a (naively) unexpected gapless quasiparticle mode, which has important implications for the physical properties of nuclear condensates.   This low-temperature mode is a collective excitation of the electrons and (condensed) ions.  The mode is a density oscillation with a linear dispersion relation, and so it is a sound-like mode which propagates at zero temperature.  However, it is not the zero sound familiar from Fermi liquids, and for lack of a better name we will call this gapless mode `half-sound'.    In \ourpreviouspaper we examined the effect of the half-sound  on the specific heat of the nuclear condensate, which helps set the rate of cooling via electromagnetic radiation.  However, the existence of the gapless half-sound mode should also affect the neutrino cooling rate, since it is not Boltzmann suppressed for any temperature $T$, in contrast to the gapped plasmon mode.   The purpose of this paper is to investigate whether the annihilation of half-sound quasiparticles into neutrinos might make an appreciable contribution to the total luminosity of He WDs, and hence to their cooling rate.

Ignoring quantum effects, at zero temperature and at the relevant densities the helium nuclei would arrange themselves, due to the Coulomb repulsion, into a lattice immersed in a neutralizing sea of electrons. At a higher temperature $T_{\textrm{melt}}$ this lattice melts. We can estimate $T_{\textrm{melt}}$ by equating the typical Coulomb energy to the temperature; the resulting estimate is $T_{\textrm{melt}} \sim  \alpha n^{1/3}$, where $n$ is the density. On the other hand, since helium ions are bosons, at low temperatures they would be expected to Bose-condense once their thermal de Broglie wavelength exceeds the inter-ion separation. The temperature at which such an ion condensate would melt can be estimated by equating the thermal wavelength of the ions to the inter-ion distance, leading to the estimate $T_{\textrm{cond}} \sim n^{2/3}/M$. This shows that at sufficiently high density, there is a range of temperatures where the ions will be found in a condensate, and not in a classical plasma or a lattice as one might have expected.  For plasmas made from heavier elements this quantum liquid region would open up only at unrealistically high densities, but for a helium plasma the relevant densities are expected to be reachable in He WDs.  The melting temperature $T_{melt}$ has been analyzed carefully in the past and is well known\cite{1975ApJ...200..306L,PhysRevA.21.2087,1993ApJ...414..695C}, at least for the parameters relevant for CO WDs. $T_{cond}$, however, is less well known. It is expected to be higher than the free gas result $T_{\textrm{cond}}^{\mathrm{free}} = 1.49\ n^{2/3}/M$ on account of the strong repulsion between ions\cite{Huang:1999zz}.  This expectation is born out by the computation in  \reference{Rosen:2010es}, which found a large enhancement over $T_{\textrm{cond}}^{\textrm{free}}$ of about an order of magnitude;  however the effects of the half-sound mode were not included in \reference{Rosen:2010es}.   With the estimates of \reference{Rosen:2010es,Gabadadze:2009jb}, the region in parameter space where a condensate is expected to occur overlaps with the relevant densities and temperatures in white dwarfs.

{\it Effective field theory.}  The coupled system of non-relativistic helium ions, electrons, and photons turns out to support three quasiparticle modes once the ions condense  \cite{Bedaque:2011hs}.  The two gapped modes are the plasmon mode and the transverse photons, with gaps set by $\omega_{p} = 4\pi Z \alpha/m_{red}$ and $m_{A} =4\pi Z^{2} \alpha v^{2}/M$ respectively, where $Z = 2$ is the charge of the nuclei, $M$ is their mass, $\alpha \approx 1/137$ is the fine structure constant, and $m_{red} = \mu_{e}M/(Z\mu_{e}+M)$ is the reduced mass of the ion-electron system, where $\mu_{e}$ is the electron chemical potential.  The chemical potential is related to the ion density $n$ via $\mu_{e}^{2} = m_{e}^{2}+k_{F}^{2}, k_{F}^{3}/3\pi^{2} = Z n$, where $m_{e}$ is the electron mass.  At typical white dwarf conditions (central density of $5\times 10^{5}\,\mathrm{g/cm^{3}}$ and  $T \lesssim   3 \times 10^{6}\,\mathrm{K} \sim m_{A}$ ),  the gapped modes are Boltzmann suppressed, and the low-energy physics can be described in terms of the interactions of the gapless half sound mode, which has the Lagrangian
\begin{align}
\label{eq:HalfSoundLag}
	\Lag_{H} = \half H\left[-\partial_{0}^{2}+c_{H}^{2}\grad^{2}	\right]H +\cdots
\end{align}
where $c_{H} = m_{A}/m_{s}$, where $m_{s} =4\alpha \mu_{e} k_{F}/\pi $ is the Debye screening mass due to the electrons, and the ellipsis denote interaction terms. Numerically, for the previously mentioned representative white dwarf, $c_{H} \sim 0.001 c$, where $c$ is the speed of light, i.e., $c_{H} \sim 300\,\mathrm{km/s}$.   

Since $H$ is is a gapless mode, the decay of $H$ to neutrinos is kinematically forbidden, and the leading contribution to the neutrino emissivity is from the annihilation reaction $HH\to \nu \bar{\nu}$.  The neutrino emissivity (energy emitted per time per volume) $Q$ can be estimated as $Q \sim (G_{F}^{2}/M^{2} ) T^{11}$.  Here the dependence on Fermi constant $G_{F}$ is the usual one for weak-interaction processes.  Meanwhile, the dependence on $M$ is a consequence of the fact that the relevant interaction term must be quadratic in $H$. The $M$ dependence then follows by noting that the H mode is derivatively coupled, so that  each power of $H$  comes with a power of spatial momentum $p$, but since the helium nuclei are non-relativistic, $p$ must enter in the combination $p^{2}/2M$.  What such naive estimates cannot tell us, however, is the dependence of $Q$ on $c_{H}$, which turns out to give a parametric enhancement.  To see that an enhancement is possible, note that the number density of $H$ quasiparticles is a function of their energy $c_{H}p$, so temperatures should enter as $(T/c_{H})$, which would yield an enhancement of order $c_{H}^{-11} \sim 10^{33}$. However, this is clearly an overestimate, since the phase space for the decay vanishes as $c\goesto0$, and a detailed calculation is necessary to determine how $c_{H}$ enters $Q$.   We will demonstrate that the correct dependence is $Q\propto c_{H}^{-7}$,  yielding a nontrivial enhancement.


{\it Matching to the SM.} To compute the neutrino emissivity (energy loss to neutrinos per time per volume), we must derive the coupling between the half-sound mode and neutrinos, which we do by a matching calculation between the Standard Model (SM) down to the half-sound EFT in Eq. \eqref{eq:HalfSoundLag}.  The relevant coupling between $H$ and neutrinos will be mediated by the neutral \Z\ bosons, since the \Z\ can decay to directly to neutrinos which can then escape the core, in contrast to the charged $W^{\pm}$ bosons that would decay into a Pauli-blocked electron and a neutrino.

\begin{figure}[tbp]
	\centering
		\includegraphics[height=3.5cm]{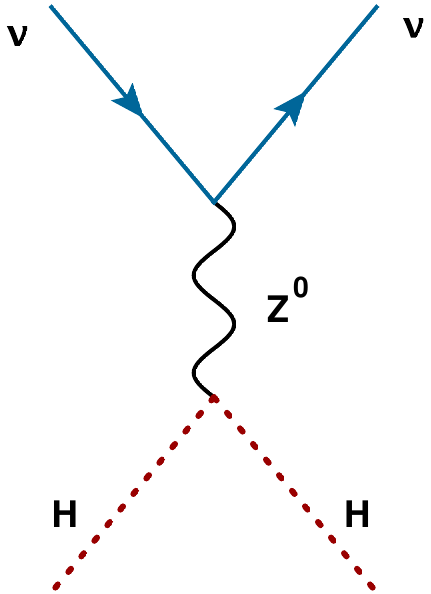}
		\caption{The annihilation of two $H$ particles into a neutrino pair.	The dashed lines are the incoming $H$ particles, and the solid lines are the outgoing neutrino/antineutrino pair.}
		\label{fig:annihilation_diagram}
\end{figure}

The $H$ is a collective mode arising from  the collective behavior of electrons and helium nuclei, and in general the $H$ field will couple to the weak sector through both its constituents.   In the SM,  \Z's coupling to the leptons is directly supplied by the Standard Model.  The relevant piece of the Lagrangian is
\begin{align}\label{eq:lag-lep}
	\Lag_{\text{\Z-lep}} = \frac{g\Z_{\mu}}{\cos\thw}&\left[\oneover{4}\bar{\nu}\gamma^{\mu}(1-\gamma_{5})\nu \right.\\
		&\left.-\oneover{4}\bar{e}\gamma^{\mu}(1-\gamma_{5})e+\sin^{2}\thw \bar{e}\gamma^{\mu}e	\right]\nn
\end{align}
where $g$ is a weak coupling constant, \thw\ is the Weinberg angle, the fields $\nu$ and $e$ are neutrinos and electrons respectively.

 
 The SM couplings of the \Z\ to quarks are described by 
\begin{align}\label{eq:lag-q}
	\Lag_{\text{\Z-q}} = \frac{g\Z_{\mu}}{\cos\thw}&\left[ \oneover{4}\bar{u}\gamma^{\mu}(1-\gamma_{5})u -\frac{2}{3}\sin^{2}\thw \bar{u}\gamma^{\mu}u \right.\\
		&\left.	 -\oneover{4}\bar{d}\gamma^{\mu}(1-\gamma_{5})d+ \oneover{3}\sin^{2}\thw\bar{d}\gamma^{\mu}d\right]\nn
\end{align}
We now need to deduce how a helium nucleus interacts with the \Z, which we do by a series of matching calculations.  First, let us consider how the \Z\ couples to the nucleon isospin doublet $N$. The \Z\ must interact as
\begin{align}\label{eq:lag-had}
	\Lag_{\text{\Z-had.}} &= \frac{g\Z_{\mu}}{\cos\thw}\left[\half \bar{N}\gamma^{\mu}T^{3}N - (g_{A}+\Delta_{s})\bar{N}\gamma^{\mu}\gamma_{5}T^{3}N\right. \nn\\
		&\left.\phantom{\half} - \sin^{2}\thw \bar{N}\gamma^{\mu}QN+\cdots\right].
\end{align}
where $g_{A}\approx 1.26$  is the nucleon axial charge, $\Delta_{s} = -0.16\pm 0.15$ the strange axial charge, the matrix $T^{3}=\tau^{3}/2$ the third component of \emph{weak}  isospin, $Q$ the electric charge in units of the fundamental charge, and the dots indicating higher dimensional terms which we will neglect. Notice that the vector coupling is not renormalized as it is a conserved current.

%

We now take the nonrelativistic reduction of \eqref{eq:lag-had} so that we can more easily read off how a condensate of nuclei should interact with \Z.  It is simple to verify that
\begin{align}\label{eq:lag-had-NR}
	\Lag_{\text{\Z-had}} &\underset{N.R.}{\longrightarrow} \frac{g}{\cos\thw}\left[\half \Z_{0} N\adj T^{3}N - (g_{A}+\Delta_{s}) \Z_{i}N\adj\sigma^{i}T^{3}N\right. \nn\\
		&\left.\phantom{\half} - \sin^{2}\thw \Z_{0}N\adj QN+\cdots\right].
\end{align}
So, we see that the \Z\ couples to the total weak isospin.  The helium nucleus has zero weak isospin, and thus does not couple to the \Z\ through the first term at all.  Because the ground state of the \Hefour\ nucleus is dominated by configurations where the total spin of the four nucleons vanishes (only $10\%-15\%$ of the wave function has non-vanishing spin \cite{Nogga:2000uu}), we expect the second term to be negligible, and so we will ignore the axial coupling.   We are left with
\begin{equation}\label{eq:lag-helium}
	\Lag_{\text{\Z-had-NR}} = -2 \frac{g\sin^{2}\thw}{\cos\thw}\Z_{0}\psi\adj\psi,
\end{equation}
where the field $\psi$ represents the \Hefour\ nuclei.  Note that this coupling is suppressed by the Weinberg angle.  In a fictitious world where $\sin\thw=0$, the \Z\ would be the gauge boson related to the third component of weak isospin and thus would not couple at all to an isosinglet source like \Hefour. 

Finally, recalling from \ourpreviouspaper that $H$ is related to $\psi$ via
\begin{equation}
	\psi = \left(	v+\sqrt{\frac{-\grad^{2}}{2M}}\frac{H}{\sqrt{2}} \right)e^{i\phi},
\end{equation}
we see from Eq.\eqref{eq:lag-helium} that the $H$-\Z\ coupling is\footnote{$H$ mixes with \Z\, but this weak force effect is negligible.}.
\begin{equation}\label{eq:lag-ZHH}
	\Lag_{\Z HH} = - \frac{g\sin^{2}\thw}{2M \cos\thw} Z_{0}(\sqrt{-\grad^{2}}H)^{2}
\end{equation}
Because the regime we are interested in includes temperatures up to roughly  $\sim 10^{ 6}K\approx 10^{-7}$GeV, which is vastly smaller than the mass of the \Z, $m_Z\approx90$GeV, we can immediately integrate out the $Z$, directly connecting the $H$ excitations with neutrinos.  A simple calculation yields
\begin{align}
\label{eq:HHnn}
	\Lag_{HH\nu\nu} &= \left(	\frac{g}{\cos\thw}	\right)^{2}\frac{\sin^{2}\thw}{8M M_{Z}^{2}}\bar{\nu}\gamma^{0}(1-\gamma_{5})\nu \left(	\sqrt{-\grad^{2}}H	\right)^{2} \nn \\
	&= \frac{\Gf\sin^{2}\thw}{\sqrt{2}M}\bar{\nu}\gamma^{0}(1-\gamma_{5})\nu \left(	\sqrt{-\grad^{2}}H	\right)^{2}
\end{align}
where in the second line we used $(g/\cos\thw)^{2} = 8 \Gf M_{Z}^{2}/\sqrt{2}$.  

{\it Neutrino emissivity.} We are now in a position to compute the neutrino emissivity $Q$, which can be written as 
\begin{equation}
\label{eq:QFormula}
	Q = \int \frac{d^{3}k}{(2\pi)^{3}}\frac{ d^{3}k'}{(2\pi)^{3}} n(k_{0})n(k'_{0})(k_{0}+k'_{0})\Gamma(k,k').
\end{equation}
where $n(k_{0})$ is the Bose-Einstein distribution $(\exp(\beta k_{0})-1)^{-1}$, $\beta = 1/T$, $k_{0},k'_{0}$ are the energies of the annihilating $H$ modes with momenta of magnitude $k, k'$, and $\Gamma(k,k')$ is the annihilation probability per time $T$ per volume $V$.  Writing the annihilation amplitude for the process of \figref{fig:annihilation_diagram}, which can be read from \eqref{eq:HHnn}, the annihilation probability per time per volume evaluates to
\begin{align}
	\Gamma_{kk'} =& \frac{\Gf^{2}\sin^{4}\thw}{M^{2}}\int \frac{\ d^{3}p}{(2\pi)^{3}}\frac{\ d^{3}p'}{(2\pi)^{3}}VT \nn\\
				&\times (2\pi)^{4}\delta^{4}(p+p'-k-k') \\
				&\times \frac{(kk')^{2}(p_{0}p_{0}'+\vec{p}\cdot\vec{p}')}{k_{0} k'_{0} p_{0} p'_{0}}\nn,
\end{align}
and we are summing over the helicities and momenta $p, p'$ of the neutrinos with energies $p_{0} \approx p, p'_{0} \approx p'$ in the final state.  The momentum integral can be evaluated in closed form, yielding
\begin{equation}
	\Gamma(k,k')= \frac{\Gf^{2}\sin^{4}\thw}{6\pi c_{H}^{2}M^{2}V} kk'\abs{k+k'}^{2}\theta\bigg[	c(k+k')-\abs{k+k'}\bigg].
\end{equation}
Plugging this result into \eqref{eq:QFormula}, the remaining integral over the momenta of the annihilating $H$ quasiparticles can be evaluated analytically in the approximation that $c_{H} \ll 1$, with the result
\begin{align}
	Q	
		&=	\frac{2048}{99\pi^{5}}\left(	\pi^{10}-93555\zeta(11)	\right)\frac{\Gf^{2}\sin^{4}\thw}{M^{2}c_{H}^{7}\beta^{11}}
\end{align}
where $\zeta$ is the usual Riemann Zeta function.  As advertised earlier, the dependence of the emissivity $Q$ on $c$ is $Q\propto c_{H}^{-7}$, a large enhancement compared to the rate one would estimate from naive dimensional analysis. 

This expression accounts for the thermal annihilation of two $H$ particles into a single neutrino pair.  Since there are three neutrino species that all couple equally the the \Z\ boson, and each of these are legitimate products of the reaction, the total power in neutrinos per unit volume emanating from a bulk of helium condensate of temperature $T$ is
\begin{align}
\label{eq:QResult}
	Q &= 		\frac{2048}{33\pi^{5}}\left(	\pi^{10}-93555\zeta(11)	\right)\frac{\Gf^{2}\sin^{4}\thw T^{11}}{M^{2}c^{7}}	\\
	  &\approx	9.5 \frac{\Gf^{2}\sin^{4}\thw T^{11}}{M^{2}c^{7}}.																\nn
\end{align}

\label{sec:Implications_to_white_dwarf_cooling}

\begin{figure}[tbp]
	\centering
		\includegraphics[width=\columnwidth 
		]{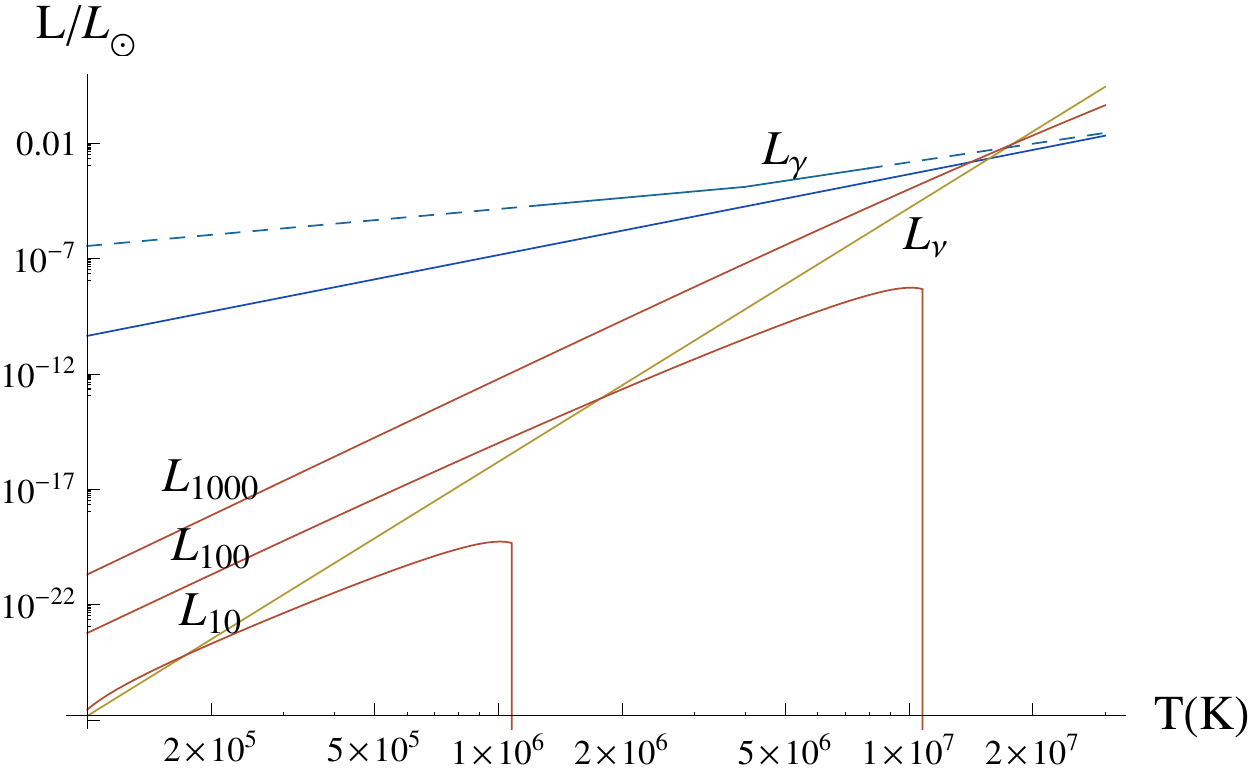}
		\caption{The  ``$L_\gamma$" (blue) lines are the electromagnetic luminosities from \eqref{eq:mestel} and from the model in \reference{Serenelli:2001vh} (the dotted line is our extrapolation). 
		The (yellow)  ``$L_\nu$" line is the neutrino luminosity for a fictitious constant-density star. The (red) $L_{10}, L_{100}$ and $L_{1000}$ are our results for a critical temperature parameter equal to $C=10,100$ and $1000$. 
		\label{fig:luminosities}}
\end{figure}

{\it Implications for white dwarf cooling.} The phenomenological importance of energy loss rate through neutrino emission depends on the ratio of the neutrino luminosity $L_{\nu}$ and the electromagnetic luminosity $L_{EM}$. The comparison between the two is complicated by the fact that the electromagnetic emission is proportional to the surface area of the star while the neutrino emission is a bulk property. In addition, the electromagnetic radiation depends on the surface temperature while the neutrino emission depends on the temperature of the  interior. Finally, the neutrino emission depends strongly on the density. All these factors make it necessary to use a model of the star structure in order  to make a meaningful comparison. Since our aim here is to understand whether the neutrino emission mechanism we identified can be competitive with  electromagnetic radiation as a cooling mechanism, we will stick to simple models of the star structure capable of giving trustable but not very accurate estimates.

The broad features of electromagnetic radiation can be understood in a model where most of the star is occupied by an isothermal, highly heat conducting degenerate core surrounded by a thin layer of non-degenerate material. Assuming that radiative heat transport dominates over convection, using a free gas equation of state for the electrons and a Kramer's opacity law for the envelope one can derive a simple relation for the star's luminosity~\cite{1952MNRAS.112..583M,phillips1994physics}:
\begin{equation}\label{eq:mestel}
\frac{L}{L_\sun} \approx \left(  \frac{T}{7\times 10^7\ K}\right)^2 \left(  \frac{M}{M_\sun}\right),
\end{equation}
where $L$ ($L_\sun$) is the star (sun) luminosity, $M$ ($M_{\sun}$) is the star (solar) mass and $T$ is the nearly constant temperature in the star interior. To be specific, we will consider a He WD with mass $M=0.406 M_\sun$. For such a star, the luminosity predicted by \eqref{eq:mestel} is shown in \figref{fig:luminosities} by the $L_\gamma$ solid (blue) line.
A more complete model for He WDs including  convection provides a similar relation between the interior temperature and the surface temperature\cite{Serenelli:2001vh}, which then gives an estimate for the electromagnetic luminosity if we assume black body radiation. This relation is also shown in \figref{fig:luminosities}, including a naive linear extrapolation (shown in dotted lines) for lower and higher temperatures than  those computed in \reference{Serenelli:2001vh}.

To estimate $L_{\nu}$ we use a simple model of the star structure given by the hydrostatic equilibrium equation:
\begin{align}\label{eq:LaneEmden}
\frac{1}{r^{2}} \frac{d}{dr} \left( \frac{r^{2}}{\rho} \frac{d P}{d r} \right) = - 4\pi G_{N} \rho
\end{align} where $\rho(r)$ is the mass density, $G_{N}$ is the Newton constant and $P = P(\rho)$ is the pressure of a (relativistic) free degenerate Fermi gas
\begin{align}
	\label{eq:P_rel}
	P &= \frac{m_{e}^{4}}{24\pi^{2}} f\left( \frac{k_F}{m_e}\right), \\  &f(x) = x(2x^{2}-3)(x^{2}+1)^{1/2}+3\sinh\inverse(x)\nonumber.
\end{align}
The solution of \eqref{eq:LaneEmden} and \eqref{eq:P_rel} provides us with the density profile of the star which we can fold into (\ref{eq:QResult}) and thence compute the $L_{\nu}$. However, (\ref{eq:QResult}) is valid only where the condensate forms.  Close to the surface of the star the density is small and so is the condensation temperature $T_{cond}$, and hence one must appropriately restrict the radial integral in $L_{\nu}$. At the center of the star the density is high and the interior temperature $T$ may be smaller than $T_{melt}$, and instead of a Bose condensed phase one may have a crystalline phase.   Consequently, to be conservative we restrict the integration both at low and large $r$, so that
\begin{align}
\label{eq:NeutrinoLuminosity}
L_{\nu}(T) = 4\pi \int_{r_{min}}^{r_{max}} dr\, r^{2} Q(T,r)
\end{align} to the region where the condensate is sure to exist. The distances $r_{min}$ and $r_{max}$ are defined by
\begin{equation}
	T= T_{cond}(r_{max}) = T_{melt}(r_{min}).	
\end{equation}
The classical crystal-melting temperature depends on the position through its dependence on the density:
\begin{equation}
	\label{eq:t_cond}
	T_{melt} \approx 120 Z^2\alpha\left( \frac{4\pi \rho}{M}\right)^{1/3}
\end{equation}
but the presence of strong quantum effects in our system may well make the true $T_{melt}$ lower.  As we will see, however, uncertainties in $r_{min}$ are not very significant  for $L_{\nu}$ since the largest contributions to $L_{\nu}$ come from large $r$. The condensation temperature also depends on the position through its dependence on the density. However, this dependence is much more uncertain. A free Bose gas would have a critical temperature  equal to
\begin{equation}
T_{cond} = \frac{C}{M^{5/3}} \left( \frac{\rho}{\zeta(3/2)}  \right)^{2/3},	
\end{equation}
with $C=1$. The repulsion between the ions raises the critical temperature by a currently unknown amount.  In \reference{Rosen:2010es} the critical temperature was found to be of the form in (\ref{eq:t_cond}) with $C\approx 9$, but this calculation did not include the effects of dynamical electrons or the massless quasiparticle. Since the value of the constant $C$ is the biggest uncertainty in our estimate we repeated our calculation for different values of $C$ covering a wide range. The results of these calculation are shown in the three (red) lines denoted by $L_{10}, L_{100}$ and $L_{1000}$ in  \figref{fig:luminosities}, corresponding respectively to the values $C=10,100$ and $1000$. For comparison we also show in  \figref{fig:luminosities} (the yellow line ``$L_\nu$") the neutrino luminosity from a fictitious `star' with a constant density equal to the average density of a $M=0.406 M_\sun$ WD star. The slope of this last curve is set by the $T^{11}$ dependence in  \eqref{eq:QResult}.  The curves $L_{10}, L_{100}$ and $L_{1000}$ represent the competition of two effects. The first is the steep dependence $Q \sim T^{11}$. The second is that the higher the temperature is, the larger the uncondensed portion of the star is, and that portion does not emit neutrinos. 
This effect is especially important because, due to the strong dependence of $Q$ on the density ($Q\sim \rho^{-7/2}$), most of the emission happens on the outer regions of the star, the first region to enter the non-condensed state as the temperature rises. The second factor dampens somewhat the rapid increase expected by the $\sim T^{11}$ behavior until a temperature higher than $T_{cond}(r=0)$, at which point the condensed phase disappears from the star and the neutrino emission comes abruptly to an end.

The curves in the \figref{fig:luminosities} have some uncertainties that should be kept in mind.  These are mainly due to the use of the $T=0$ dispersion relation for $H$, and to the reliance of our computation on the low-momentum limit of the dispersion relations.   The latter point is an issue because away from small $p$, the dispersion relation of $H$ is not linear, with the deviations becoming important for modes with energies $T \sim p_{0} \gtrsim m_{A}$ (recall from \ourpreviouspaper that $m_{A}$ ranges between $2.1\cdot 10^{6}$K and $4.6\cdot 10^{6}$K as $a_{0}/l$ goes from 30 to 50, where $a_{0}$ is the Bohr radius and $l$ the interparticle spacing).  Hence we can only really trust our results for $L$ where $T < m_{A}$.  The use of the $T=0$ dispersion relation for $H$ may be problematic, since even  when $T\ll T_{c}$ near the center of the star, close to the surface one will probe regions where $T\sim T_{c}$ thanks to the decrease in density.  In general, one would expect that the effect of a finite temperature would be to lower $m_{A}$ by lowering the size of the nuclear condensate, which would then lower $c_{H}$, increasing the luminosity. If the phase transition to the uncondensed phase is first order, as suggested by \cite{Rosen:2010es}, then the drop in $m_{A}$ close to $T_{c}$ would not be parametrically large, and such effects should not affect our conclusions qualitatively.  If instead the phase transition is second-order, there is an enhancement of the luminosity at points where $v$ is small, but that enhancement should be cut off by the fact that the dispersion relation will be linear over a much smaller range of momentum.  A better assessment of these finite-temperature effects must wait until the thermodynamics of this transition is better understood.



{\it Conclusions.} The results shown in  \figref{fig:luminosities} indicates that neutrino emission from the condensed phase can be an important sink of energy in a He WD if and only if the critical temperature for condensation is quite high. More systematic calculations of the critical temperature and of the finite-$T$ corrections to the dispersion relations are currently under way \cite{preparation}.   In the case the critical temperature turns out to be high it would be important to include the neutrino emission from the condensed phase in a realistic cooling code. Only then could one ascertain whether the cooling of He WD can be a smoking gun signature for the existence of a quantum liquid phase inside white dwarfs.

{\it Acknowledgements} P.~F~B. and E.~B. are supported by the U.S. Dept. of Energy under grant \#DE-DG02-93ER-40762. E.~B. also thanks Jefferson Science Associates for support under the JSA/JLab Graduate Fellowship program, and A~.C. thanks the STFC for support through the theory group grant at DAMTP.

\bibliography{nuclear_liquids,astro_refs}
\end{document}